\documentclass{PoS}

\title{Electroweak Penguin Hunting Through $B\to\pi\pi,\pi K$ and Rare $K$ and $B$ Decays}

\ShortTitle{An Analysis of the $B\to\pi K$ Puzzle and its Relation to Rare Decays}

\author{Andrzej J. Buras
\hspace*{7.8cm}\parbox{3cm}{\rm\normalsize 
hep-ph/0512059\\\mbox{CERN-PH-TH/2005-242}\\TUM-612/05\\
MPP-2005-157\\[-1.5cm]}\\
        TU M\"unchen, Germany\\
        E-mail: \email{Andrzej.Buras@ph.tum.de}}
        
\author{\speaker{Robert Fleischer}\\ %\thanks{}\\
        CERN, Switzerland\\
        E-mail: \email{Robert.Fleischer@cern.ch}}

\author{Stefan Recksiegel\\
        TU M\"unchen, Germany\\
        E-mail: \email{Stefan.Recksiegel@ph.tum.de}}
    
\author{Felix Schwab\\
        TU M\"unchen and Max-Planck-Institut f\"ur Physik, Germany\\
        E-mail: \email{Felix.Schwab@ph.tum.de}}

\abstract{The $B\to\pi K$ decays with significant electroweak penguin contributions 
show a puzzling pattern. We explore this ``$B\to\pi K$ puzzle" through a 
systematic strategy. The starting point, which is essentially unaffected by 
electroweak penguins, is the determination of the angle $\gamma$ of the 
unitarity triangle through the CP-violating $B^0_d\to\pi^+\pi^-$, $B^0_d\to\pi^-K^+$ 
asymmetries, yielding $\gamma=(73.9^{+5.8}_{-6.5})^\circ$, and the extraction of 
hadronic parameters through the measured $B\to\pi\pi$
branching ratios. Using arguments related to the $SU(3)$ flavour symmetry, we 
convert the hadronic $B\to\pi\pi$ parameters into their $B\to\pi K$ counterparts, 
allowing us to predict the $B\to\pi K$ observables in the Standard Model. We find 
agreement with the data for those quantities that are only marginally affected by
electroweak penguins, while this is not the case for the observables with sizeable
electroweak penguin contributions. Since we 
may also perform a couple of internal consistency checks of our working assumptions, 
which are nicely satisfied for the current data, and find a small sensitivity of our 
results to large non-factorizable $SU(3)$-breaking corrections, the ``$B\to\pi K$"
puzzle may be due to new physics in the electroweak penguin sector. We show that
it can indeed be resolved through such a kind of new physics with a large CP-violating
phase. Further insights into the electroweak penguins are provided by the 
$B^+\to\pi^0K^+$ and $B_d^0\to\pi^0K_{\rm S}$ CP asymmetries, and 
in particular through correlations with various rare $K$ and $B$ decays. 
} 

\FullConference{International Europhysics Conference on High Energy
Physics\\
		 July 21st - 27th 2005\\
		 Lisboa, Portugal}

\begin{document}
\section{Introduction}
Since the observation of the decay $B^0_d\to\pi^0K^0$ by the CLEO collaboration
in 2000 with a remarkably prominent rate, we are faced with a possible discrepancy 
with the picture of the Standard Model (SM). This ``$B\to\pi K$ puzzle" is still
present in the most recent $B$-factory data, and has recently received a lot of
attention. In \cite{BFRS}, where also a comprehensive list of the relevant literature
can be found, we developed a strategy to explore this exciting topic in a systematic
manner. The starting point is an analysis of the $B\to\pi\pi$ system, where the
data can be accommodated in the SM through large non-factorizable effects. In
particular, the $B\to\pi\pi$ decays allow us to extract a set of hadronic parameters 
with the help of the isospin symmetry of strong interactions. Using then the $SU(3)$
flavour symmetry and neglecting certain exchange and penguin annihilation topologies,
we can convert the hadronic $B\to\pi\pi$ parameters in their $B\to\pi K$ counterparts,
allowing us to predict all $B\to\pi K$ observables in the SM. We find agreement for
those decays that are only marginally affected by (colour-suppressed) electroweak (EW)
penguins. On the other hand, the SM predictions of the $B\to\pi K$ observables which 
are significantly affected by (colour-allowed) EW penguins do not agree with the
data, thereby reflecting the $B\to\pi K$ puzzle. Moreover, we can perform internal
consistency checks of our working assumptions, which work well within the current
uncertainties, and find that our results are very stable under large non-factorizable
$SU(3)$-breaking corrections. In view of these features, new physics (NP) in the
EW penguin sector may be at the origin of the $B\to\pi K$ puzzle. In fact, it can
be resolved through a modification of the EW penguin parameters, involving in 
particular a large CP-violating NP phase that vanishes in the SM. The implications
of this kind of NP on rare $K$ and $B$ decays are then investigated in the final
step of our strategy. 

The numerical results presented below refer to our very recent analysis 
\cite{BFRS-05}. A somewhat surprising development of this summer 
is a new world average for $(\sin 2\beta)_{\psi K_{\rm {S}}}$, which went 
down by about 1$\sigma$. The picture in the $\bar\rho$--$\bar\eta$ plane 
with the fits of the unitarity triangle (UT) is now no longer ``perfect", 
which may indicate NP in $B^0_d$--$\bar B^0_d$ mixing. Consequently,
we use the CP asymmetries of the $B^0_d\to\pi^+\pi^-$, $B^0_d\to\pi^-K^+$ 
system to determine the ``true" value of the UT angle $\gamma$, yielding
\begin{equation}\label{gamma-det}
\gamma=(73.9^{+5.8}_{-6.5})^\circ,
\end{equation}
and use it as an input for our $B\to\pi\pi,\pi K$ analysis. If we complement 
(\ref{gamma-det}) with $|V_{ub}/V_{cb}|$ (semi-lept.\ $B$ decays), we may 
also extract the ``true" value of $\beta$. We obtain 
$\beta=(25.8\pm 1.3)^\circ$, which would correspond to a 
NP phase $\phi_d^{\rm NP}=-(8.2\pm 3.5)^\circ$ in $B^0_d$--$\bar B^0_d$ mixing,
in accordance with~\cite{UTfit}.

\section{The $B\to\pi\pi$ System}
The starting point of our $B\to\pi\pi$ study is given by the following ratios:
\begin{eqnarray}
R_{+-}^{\pi\pi}&\equiv&2\left[\frac{\mbox{BR}(B^+\to\pi^+\pi^0)
+\mbox{BR}(B^-\to\pi^-\pi^0)}{\mbox{BR}(B_d^0\to\pi^+\pi^-)
+\mbox{BR}(\bar B_d^0\to\pi^+\pi^-)}\right]=F_1(d,\theta,x,\Delta;\gamma)
\stackrel{\rm exp}{=}2.04\pm0.28
\label{Rpm-def}\\
R_{00}^{\pi\pi}&\equiv&2\left[\frac{\mbox{BR}(B_d^0\to\pi^0\pi^0)+
\mbox{BR}(\bar B_d^0\to\pi^0\pi^0)}{\mbox{BR}(B_d^0\to\pi^+\pi^-)+
\mbox{BR}(\bar B_d^0\to\pi^+\pi^-)}\right]=F_2(d,\theta,x,\Delta;\gamma)
\stackrel{\rm exp}{=}0.58\pm0.13.
\end{eqnarray}
Here we have used the isospin symmetry of strong interactions to express
these observables in terms of $\gamma$ and the hadronic parameters $de^{i\theta}$,
$xe^{i\Delta}$ that were introduced in \cite{BFRS}, and have also given the
current experimental numbers. Moreover, we have the CP asymmetries
\begin{eqnarray}
{\cal A}_{\rm CP}^{\rm dir}(B_d\to \pi^+\pi^-)&=&
G_1(d,\theta;\gamma)\stackrel{\rm exp}{=}-0.37\pm0.10 \\
{\cal A}_{\rm CP}^{\rm mix}(B_d\to \pi^+\pi^-)&=&
G_2(d,\theta;\gamma,\phi_d)\stackrel{\rm exp}{=}+0.50\pm0.12
\end{eqnarray}
at our disposal, where $\phi_d\stackrel{\rm exp}{=}(43.4^{+2.6}_{-2.4})^\circ$ is the 
$B^0_d$--$\bar B^0_d$ mixing phase; its numerical value follows from
the data for CP violation in $B_d\to J/\psi K_{\rm S}$. If we use the value of
$\gamma$ in (\ref{gamma-det}), we are in a position to determine the
hadronic parameters characterizing the $B\to\pi\pi$ system, with the
following results:
\begin{equation}\label{d-theta-x-Delta}
d=0.52^{+0.09}_{-0.09}, \quad 
\theta=(146^{+7.0}_{-7.2})^\circ, \qquad
x=0.96^{+0.13}_{-0.14}, \quad 
\Delta=-(53^{+18}_{-26})^\circ.
\end{equation}
These numbers, which exhibit large non-factorizable effects, are in excellent 
agreement with our previous analysis \cite{BFRS-up}. 
Let us stress that we have also included EW penguin effects in 
(\ref{d-theta-x-Delta}) through the isospin symmetry, although these topologies 
have a minor impact on the $B\to\pi\pi$ decays. 

Finally, we may predict the CP asymmetries of the decay $B_d\to\pi^0\pi^0$, 
where we obtain 
\begin{equation}\label{ACP-Bdpi0pi0-pred}
{\cal A}_{\rm CP}^{\rm dir}(B_d\to \pi^0\pi^0)=-0.30^{+0.48}_{-0.26}
\,\stackrel{\rm exp}{=}\, -0.28^{+0.40}_{-0.39}, \quad
{\cal A}_{\rm CP}^{\rm mix}(B_d\to \pi^0\pi^0)=-0.87^{+0.29}_{-0.19}.
\end{equation}
Here we have also included the experimental value for the direct CP 
asymmetry \cite{HFAG}. Although no stringent test of our predictions is
currently provided, the indicated agreement is very encouraging.

\section{The $B\to\pi K$ System}
If we use now the $SU(3)$ flavour symmetry and neglect exchange and 
penguin annihilation topologies, we can convert the hadronic parameters in
(\ref{d-theta-x-Delta}) into their $B\to\pi K$ counterparts, allowing us to predict 
the $B\to\pi K$ observables in the SM. Moreover, a couple of internal
consistency checks of these working assumptions can be performed, which
are nicely fulfilled by the current data, and the sensitivity of our SM predictions
on large non-factorizable $SU(3)$-breaking effects turns out to be surprisingly
small \cite{BFRS-05}. Consequently, no anomaly is indicated in this sector. 

\begin{figure}
\begin{center}
\includegraphics[width=7.0cm]{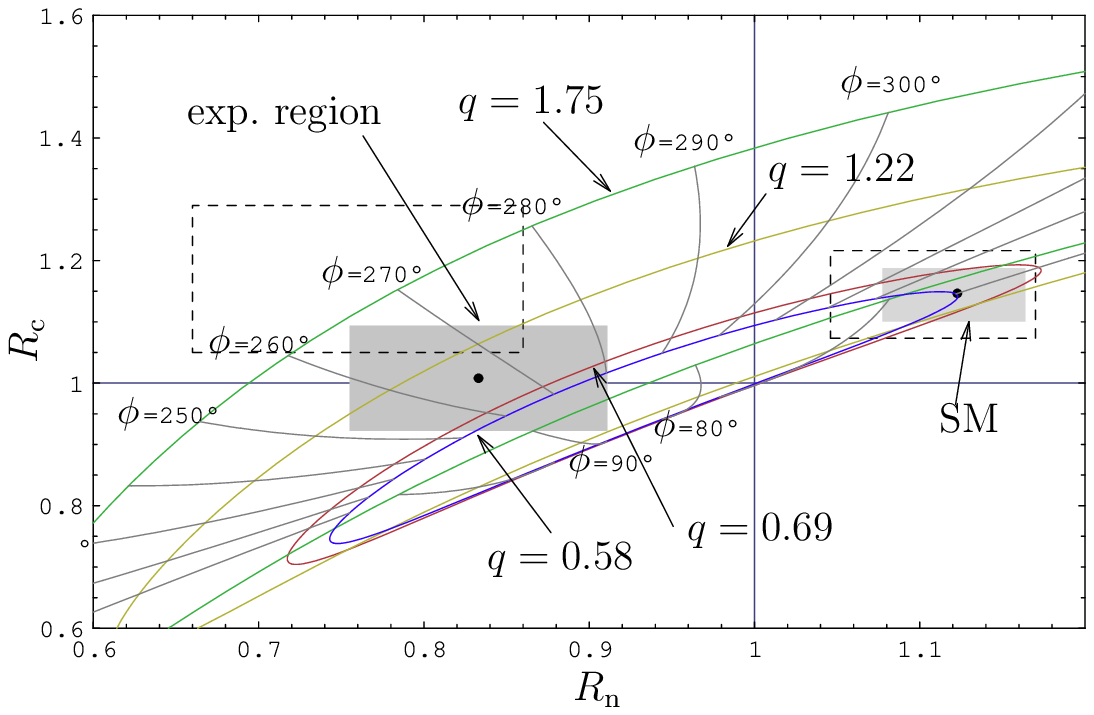}
\end{center}
\vspace*{-0.7truecm}
\caption{The situation in the $R_{\rm n}$--$R_{\rm c}$ plane, as discussed
in the text.}\label{fig:RnRc}
\end{figure}

In the case of the $B^0_d\to\pi^-K^+$, $B^+\to\pi^+K^0$ system, where 
EW penguins have a minor impact, we obtain a SM picture in accordance
with the data. In order to analyse the decays $B^+\to\pi^0K^+$ and 
$B^0_d\to\pi^0K^0$, which are significantly affected by EW penguins,
it is useful to introduce
\begin{eqnarray}
R_{\rm c}&\equiv&2\left[\frac{\mbox{BR}(B^+\to\pi^0K^+)+
\mbox{BR}(B^-\to\pi^0K^-)}{\mbox{BR}(B^+\to\pi^+ K^0)+
\mbox{BR}(B^-\to\pi^- \bar K^0)}\right] \,\stackrel{\rm exp}{=}\, 1.01\pm 0.09\\
R_{\rm n}&\equiv&\frac{1}{2}\left[\frac{\mbox{BR}(B_d^0\to\pi^- K^+)+
\mbox{BR}(\bar B_d^0\to\pi^+ K^-)}{\mbox{BR}(B_d^0\to\pi^0K^0)+
\mbox{BR}(\bar B_d^0\to\pi^0\bar K^0)}\right] \,\stackrel{\rm exp}{=}\, 0.83\pm0.08.
\end{eqnarray}
The EW penguin effects are described by a parameter $q$, which measures
the strength of the EW penguins with respect to tree-diagram-like topologies,
and a CP-violating phase $\phi$. In the SM, this phase vanishes,
and $q$ can be calculated with the help of the $SU(3)$ flavour symmetry,
yielding a value of 0.58. The situation can transparently be discussed in the 
$R_{\rm n}$--$R_{\rm c}$ plane, as shown in Fig.~\ref{fig:RnRc}:
the shaded areas indicate our SM prediction and the experimental range, 
the lines show the theory predictions for the central values of the hadronic 
parameters and various values of $q$ with $\phi\in[0^\circ,360^\circ]$;
the dashed rectangles represent the SM predictions and experimental ranges 
at the time of our original analysis \cite{BFRS}.  Although the central values of 
$R_{\rm n}$ and $R_{\rm c}$ have slightly moved towards each other, the 
puzzle is as prominent as ever. 
The experimental region can now be reached without an enhancement of $q$, 
but a large CP-violating phase $\phi$ of the order of $-90^\circ$ is
still required, although $\phi\sim+90^\circ$ can also bring us rather close
to the experimental range of $R_{\rm n}$ and $R_{\rm c}$. 

We may also predict the CP-violating asymmetries of the 
$B^\pm\to\pi^0 K^\pm$ and $B_d\to\pi^0K_{\rm S}$ decays both in the SM 
and in our NP scenario. In particular the mixing-induced CP asymmetry
of the latter decay has recently received a lot of attention, as the current
$B$-factory data yield a value of $-0.38\pm 0.26$ for 
$\Delta S \equiv (\sin2\beta)_{\pi^0K_{\rm S}}-
(\sin2\beta)_{\psi K_{\rm S}}$. We predict this difference to be {\it positive}
in the SM, and in the ballpark of $0.10$--$0.15$ \cite{BFRS-05}. Interestingly,
the best values for $(q,\phi)$ that are implied by $R_{\rm n,c}$ make the 
disagreement of $\Delta S$ with the data even larger than in the SM. However, 
also values of $(q,\phi)$ can be found for which $\Delta S$ could be smaller than
in the SM or even reverse the sign \cite{BFRS-05}.

\section{Rare $K$ and $B$ Decays}
An important feature of our strategy is a connection between the
$B\to\pi\pi,\pi K$ modes and rare decays of the kind $K^+\to\pi^+\nu\bar\nu$,
$K_{\rm L}\to\pi^0\nu\bar\nu$, $K_{\rm L}\to\pi^0\ell^+\ell^-$, $B\to X_s\nu \bar\nu$ 
and $B_{s,d}\to\mu^+\mu^-$. If we assume that the dominant NP contributions 
enter through the $Z^0$-penguin function $C$ and make the renormalization-group 
evolution from scales ${\cal O}(M_W,m_t)$ down to ${\cal O}(m_b)$, we can 
directly explore the interplay of the modified EW penguin sector with these rare 
decays, which shows that we may encounter sizeable effects, in particular
in the $K\to\pi\nu\bar\nu$ system. In \cite{BFRS-05}, we point out that the most 
recent $B$-factory constraints for rare decays have interesting new implications, 
and discuss  a few  future scenarios. We look forward to confronting our strategy 
with more accurate data!


\begin{thebibliography}{99}
%
%
%
\bibitem{BFRS}A.~J.~Buras, R.~Fleischer, S.~Recksiegel and F.~Schwab,
  %``B $\to$ pi pi, new physics in B $\to$ pi K and implications for rare K and
  %B decays,''
  Phys.\ Rev.\ Lett.\  {\bf 92} (2004) 101804
  [arXiv:hep-ph/0312259];
  %%CITATION = HEP-PH 0312259;%%
%``Anatomy of prominent B and K decays and signatures of CP-violating new
  %physics in the electroweak penguin sector,''
  Nucl.\ Phys.\ B {\bf 697} (2004) 133
  [arXiv:hep-ph/0402112].
  %%CITATION = HEP-PH 0402112;%%

\bibitem{BFRS-05}A.~J.~Buras, R.~Fleischer, S.~Recksiegel and F.~Schwab,
  %``New Aspects of B $\to$ pi pi, pi K and their Implications for Rare
  %Decays,''
  arXiv:hep-ph/0512032.
  %%CITATION = HEP-PH 0512032;%%

\bibitem{UTfit}M.~Bona {\it et al.}\  [UTfit Collaboration],
  %``The UTfit collaboration report on the status of the unitarity triangle
  %beyond the standard model. I: Model-independent analysis and minimal flavour
  %violation,''
  arXiv:hep-ph/0509219.
  %%CITATION = HEP-PH 0509219;%%

\bibitem{BFRS-up}A.~J.~Buras, R.~Fleischer, S.~Recksiegel and F.~Schwab,
  %``The B $\to$ pi pi, pi K puzzles in the light of new data: Implications for
  %the standard model, new physics and rare decays,''
  Acta Phys.\ Polon.\ B {\bf 36} (2005) 2015
  [arXiv:hep-ph/0410407].
  %%CITATION = HEP-PH 0410407;%%
  
\bibitem{HFAG}Heavy Flavour Averaging Group: 
{\tt http://www.slac.stanford.edu/xorg/hfag/}.


%
%
%
\end{thebibliography}
\end{document}